\documentclass[conference]{IEEEtran}
\IEEEoverridecommandlockouts
\usepackage {tikz}
\usepackage{latexsym}
\usepackage{cite}
\usepackage{amsmath}
\usepackage{float}
\usepackage{graphicx}
\usepackage{caption}
\usepackage{subcaption}
\usepackage{siunitx}
\usepackage{etoolbox}
\usepackage{enumitem}

\ifCLASSINFOpdf
\else
\fi
 \makeatletter
 \patchcmd{\@maketitle}
   {\addvspace{0.5\baselineskip}\egroup}
   {\addvspace{-2\baselineskip}\egroup}
   {}
   {}
 \makeatother
\begin{document}
%
\title{Seamless Handover in IP over ICN Networks: a Coding Approach\vspace{-0.5em}}

\author{\IEEEauthorblockN{Mohammed Al-Khalidi, Nikolaos Thomos, Martin J. Reed, Mays F. AL-Naday}%
\IEEEauthorblockA{
University of Essex\\
Colchester, CO4 3SQ, UK\\
Email:\{mshawk,nthomos,mjreed,mfhaln\}\\
@essex.ac.uk}%
\and%
\IEEEauthorblockN{Dirk Trossen}%
\IEEEauthorblockA{InterDigital Europe, Ltd.\\
London, EC2A 3QR, UK\\
Email: Dirk.Trossen\\
@InterDigital.com}%
\thanks{This work was carried out within the project POINT, which has received funding from the European Union's Horizon 2020 research and innovation programme under grant agreement No 643990.}}

\maketitle

\begin{abstract}

Seamless connectivity plays a key role in realizing QoS-based delivery in mobile networks. However, current handover mechanisms hinder the ability to meet this target, due to the high ratio of handover failures, packet loss and service interruption. These challenges are further magnified in Heterogeneous Cellular Networks (HCN) such as Advanced Long Term Evolution (LTE-Advanced) and LTE in unlicensed spectrum (LTE-LAA), due to the variation in handover requirements. Although mechanisms, such as Fast Handover for Proxy Mobile IPv6 (PFMIPv6), attempt to tackle these issues; they come at a high cost with sub-optimal outcomes. This primarily stems from various limitations of existing IP core networks. In this paper we propose a novel handover solution for mobile networks, exploiting the advantages of a revolutionary IP over Information-Centric Networking (IP-over-ICN) architecture in supporting flexible service provisioning through anycast and multicast, combined with the advantages of random linear coding techniques in eliminating the need for retransmissions. Our solution allows coded traffic to be disseminated in a multicast fashion during handover phase from source directly to the destination(s), without the need for an intermediate anchor as in exiting solutions; thereby, overcoming packet loss and handover failures, while reducing overall delivery cost. We evaluate our approach with an analytical and simulation model showing significant cost reduction compared to PFMIPv6.

\end{abstract}


%
\IEEEpeerreviewmaketitle

\section{Introduction}

Heterogeneous Cellular Networks (HCNs) \cite{andrews2013seven} have been deployed to accommodate the rapid growth in mobile services and data traffic~\cite{ericsson2012ericsson}. A HCN typically consists of conventional eNodeBs and Small-cell Base Stations (SBSs)\cite{andrews2013seven}, with the latter being deployed in various numbers within different areas (i.e. variable SBS density). SBSs often have different capabilities, reflected in supporting various size coverage areas, which requires different handover configuration parameters; specifically, the Time-To-Trigger (TTT) a handover. Tuning the TTT is a critical task for current HCNs due to the existing handover solutions that consider initiating handover to a single destination, such as: X2 interface handover in LTE networks~\cite{li2016handover,xenakis2014mobility}; and, Fast Handover for Proxy Mobile IPv6 (PFMIPv6), described in the IETF standard~\cite{yokota2010fast}.

\begin{figure}[t]
 	\centering
	\includegraphics[width=0.7\columnwidth,height=1.2in]{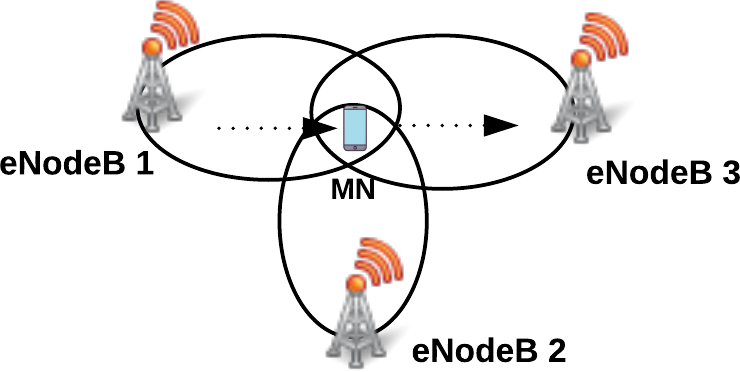}
 	\caption{Handover example for a Mobile Node.}
	\vspace{-5mm}
 	\label{HOFailure}
 \end{figure}

TTT adjustment in a single destination solution may have considerable drawbacks on the network performance and QoS; because, it may trigger a ``too-early'' handover, ``too-late'' handover, or handover to an unprepared cell, resulting in handover failure. Moreover, it may cause a ``ping-pong'' handover among neighbouring cells~\cite{lopez2012theoretical}. The significance of these issues increases as the SBS density also increases, due to the higher number of handovers. The TTT adjustment in HCNs for optimum handover success rates is investigated in\cite{vasudeva2014analysis,lopez2012mobility,chen2015theoretical}, where it has been shown that a rate of handover failure of at least 20\% of total handovers is inevitable in high mobility environments. Moreover, to achieve an acceptable balance between handover failure and ping-pong handover rates, the failure rate should be between 30\% to 60\% of total handovers depending on the mobile node (MN) velocity. An example that illustrates the failure scenarios is depicted in Fig.~\ref{HOFailure}, where in the case of ``too-late'' handover, a connection failure may occur in the source cell (eNodeB 1) and the MN may try to re-establish the radio link in a different cell i.e. (eNodeB 2). HO failure could also happen due to ``too-early'' handover, where the MN may experience a connection failure in the target cell (eNodeB 3). Hence, a MN may try to re-establish the radio link in the original source cell (eNodeB 1) which would have deleted all contexts related to that MN on handover completion. Another reason of failure is when the handover is to an unprepared cell which has not received any context related to that MN.

Handover failures could have a detrimental effect in users' Quality of Services (QoS). A potential solution could be to prepare handover to multiple destinations instead of single. This solution, however, imposes an infeasible cost in the existing HCN, IP-based, core. This is because preparing a single destination to handover requires a central topological anchor point in the network core that keeps track of a node's movement and its IP address. The anchor facilitates the handover process through a form of tunnelling between the source Base Station and itself; as well as between itself and the destination Base Station, over which traffic is rerouted from the source to the destination while maintaining the same IP address of the MN~\cite{yokota2010fast,lee2013comparative}. An additional, direct, tunnel is also established between the source and destination of the handover process, to transfer the contextual information and buffered packets of the MN. Following the same approach, in preparing multiple destinations for handover, would mean to establish (and tear down) a number of tunnels that is equal to the number of possible destinations. From the above, it is obvious that while the tunnelling cost is bearable when preparing a single destination, it is simply infeasible for multiple destinations. To overcome the above mentioned issues and reduce the handover preparation cost, we adopt an alternative network architecture that connects IP edge networks over an Information-Centric Networking (ICN) core, such as that of~\cite{7194109}. The adopted architecture substantially simplifies the requirements to prepare multiple destinations for handover, by exploiting the advantages of relaxed coupling between ICN names and nodes, as well as the stateless multicast source routing mechanism. Consequently, it eliminates the need for tunnelling to a central anchor point back in the core. Instead, our solution establishes a direct source-routed path from the source of the traffic to the new handover destination.

The use of ICN as the network core introduces new possibilities to overcome packet loss and high latency during handover, by utilizing random linear coding (RLC) techniques such as networking coding \cite{AhlswedeTIT00,ThomosTCSVT2010} and fountain codes \cite{shokrollahi2006raptor,Luby02}. These codes have been proposed as an efficient mean to combat loss, increase the throughput and decrease the experienced delays in both wired and wireless communication. RLC codes have a random coding structure and can generate a potentially rateless stream of coded packets from a given set of packets. These codes have probabilistic decoding guarantees and receivers can recover the source packets once they have collected a full rank set of RLC encoded packets. The benefits of RLC based solutions in ICN have been illustrated in \cite{AnastasiadesICC15,SaltarinINFOCOM16}. Specifically, network coding can significantly decrease the content retrieval delay, improve the resilience to packet loss and take advantage of multi-path communication \cite{SaltarinINFOCOM16}, while Raptor codes can be efficient in mobile scenarios \cite{AnastasiadesICC15}.

In this paper, we propose a seamless handover solution that exploits the advantages of stateless multicast source-routing and RLC codes in eliminating handover failure and reducing delivery cost. Our solution facilitates handover with three phases: preparation, execution and completion; whereby, all possible destinations are prepared for handover with feasible cost, before handover execution takes place. Thereby, eliminating scenarios of ``too-early'', ``too-late'' handovers as well as handover to unprepared cell. Furthermore, our solution utilizes RLC coding techniques to overcome packet loss and reduce end-to-end latency during handover. We evaluate our solution by modelling the imposed cost of the handover mechanism and compare it to that of the PFMIPv6 counterpart. Our evaluation shows that in addition to eliminating handover failures and packet loss, there is a considerable reduction in the cost of traffic delivery both during and outside handover periods. Furthermore, we illustrate seamless session continuity in high mobility environments with lower cost than that of PFMIPv6.

The rest of the paper is structured as follows, in Section 2 we explain the proposed seamless handover solution in detail, while in Section 3 we present the cost analysis for all schemes under comparison. In Section 4, we discuss the simulation environment and the obtained results. Finally, conclusions are drawn in Section 5.

 \begin{figure}[t]
 	\centering
	\includegraphics[width=\columnwidth]{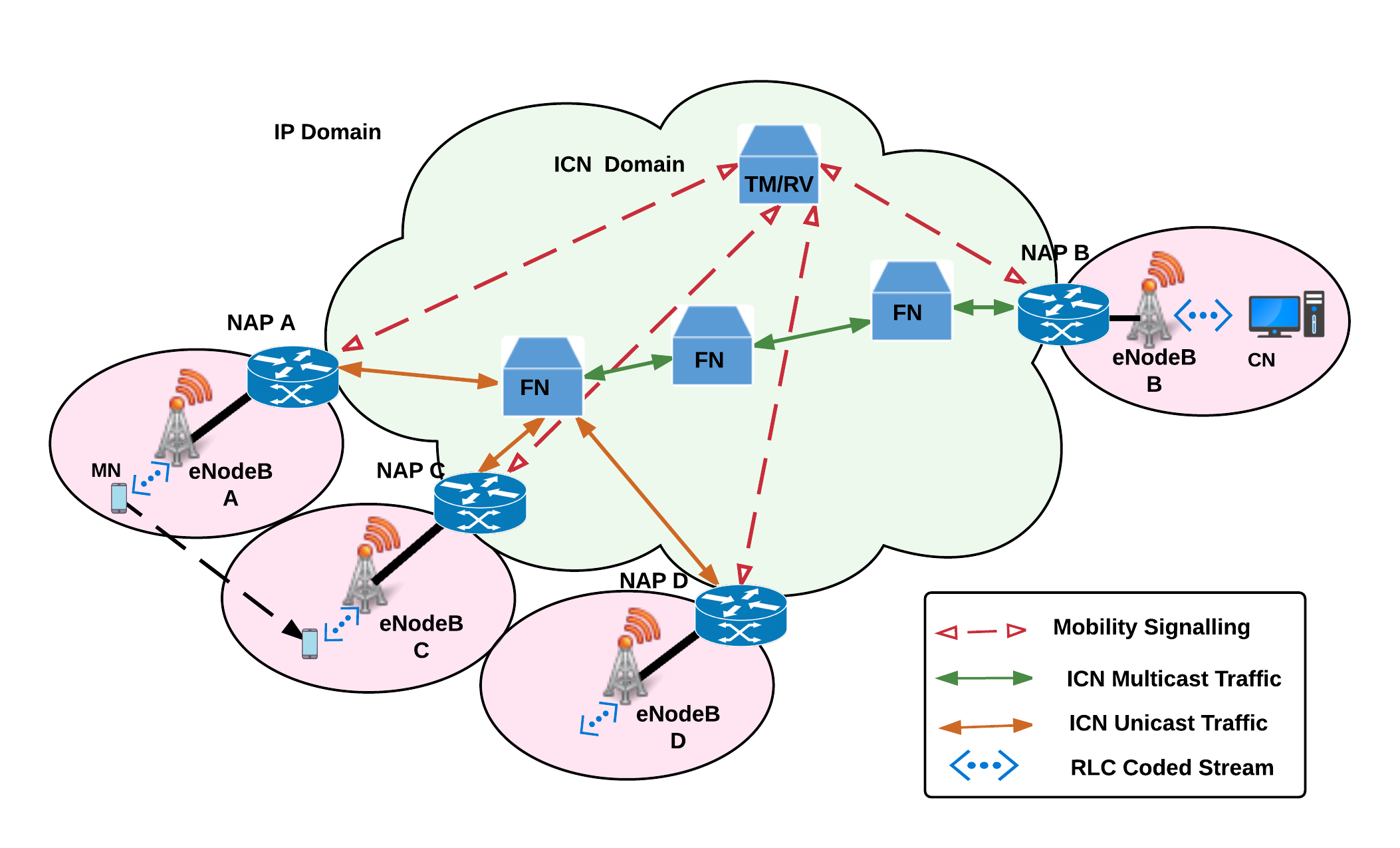}
 	\caption{Seamless Handover in IP over ICN.}
	\vspace{-5mm}
 	\label{Fast HO}
 \end{figure}

  \begin{figure*}
  	\centering
  	\includegraphics[width=5in,height=4in]{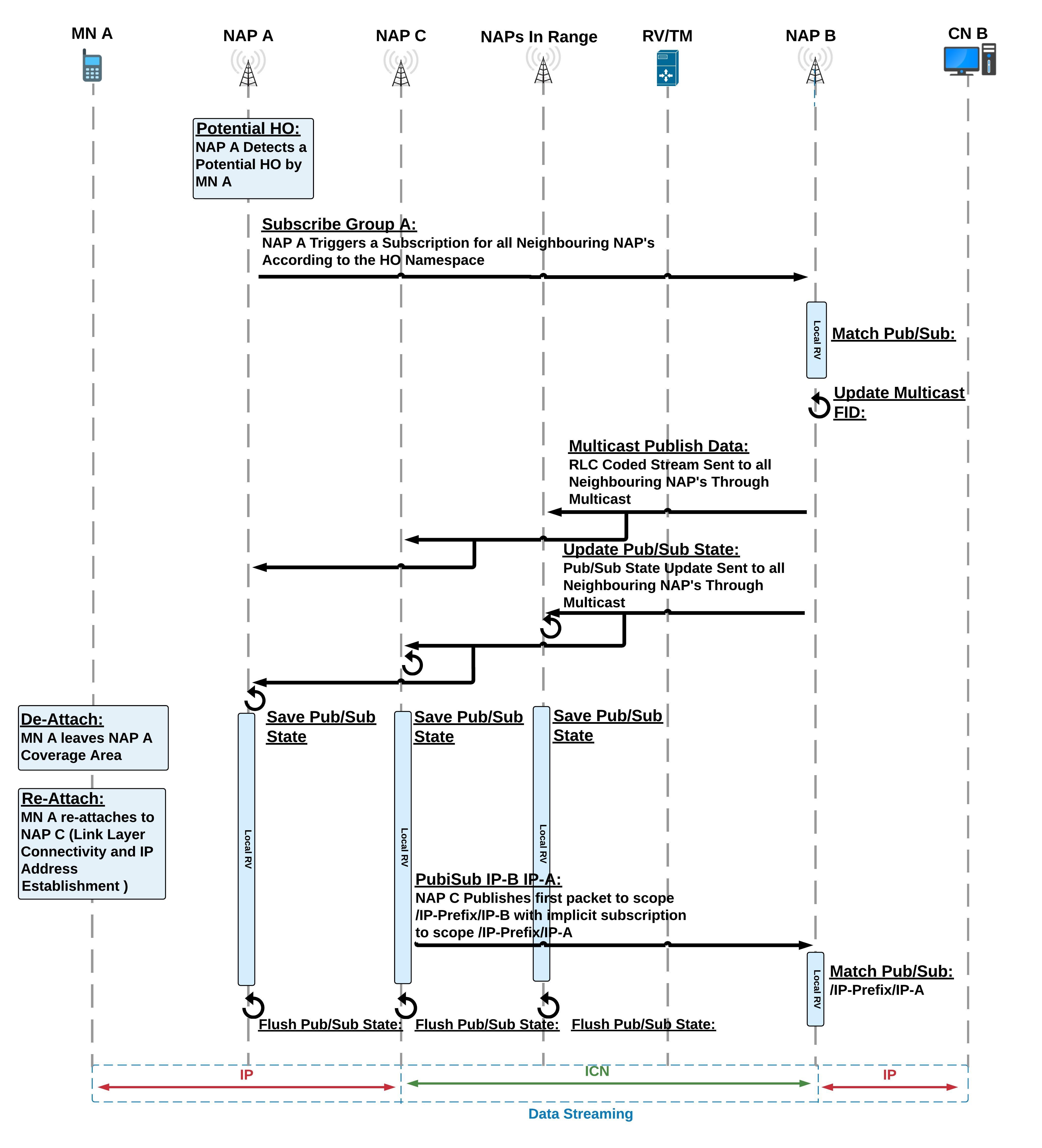}
  	\caption{Handover Sequence Diagram.}
	\vspace{-5mm}
  	\label{MBB2}
  \end{figure*} 

\section{Seamless Handover for IP over ICN Networks}
 
In this paper, we consider an IP over ICN architecture that follows a gateway based approach as shown in Fig. \ref{Fast HO}. In this architecture, the first link from the user device to the network is based on IP-based protocols, while the Network Attachment Point (NAP) serves as an entry point to the ICN network and maps the chosen protocol abstraction to ICN \cite{7194109,trossen2012designing}. The ICN uses a rendezvous (RV) function to map Pub/Sub requests to information identifiers (ICN ``names'') and a topology manager (TM) to manage forwarding in the ICN. We focus on an IP over ICN solution to cope with the sub-optimal handling of MN mobility in the core of existing cellular networks where a central anchor point is used to route and encapsulate user plane traffic towards MN's. This problem can be overcome with the employment of an Information Centric Core that naturally supports multicast and end-to-end shortest path user plane traffic routing rather than central anchored routing. The ICN core is invisible to the MN that communicates using the IPv4/IPv6 stack through a network attachment point (NAP) collocated at eNodeB. In this architecture, IP simply becomes a service enabled through the ICN core \cite{7194109}.

To facilitate seamless handover in IP over ICN networks we utilize a specific namespace for handover candidates, placed under a unique $/$root identifier. Under this root \emph{scope}\footnote{Here \emph{scope} refers to a grouping of information items in the ICN namespace, scopes may hierarchically contain other scopes.}, each NAP has an ID represented as an ICN scope under which fall the appropriate ICN names of its neighbouring NAP IDs. Hence, the scope $/root/NAP\_A$ will contain the IDs of all NAP's that are currently neighbours to the NAP represented by the identifier $NAP\_A$. A hashing scheme can be used to derive appropriate ICN identifiers from administratively assigned NAP IDs.
First, we will assume that two devices MN A and MN B are communicating and attached to NAPs A and B respectively. To do this, NAP A will have \emph{subscribed} to receive packets destined to the IP address of MN A, and likewise NAP B for MN B; then for MN A to send traffic to MN B, NAP A will have \emph{published} data to NAP B and vice versa for MN B sending to MN A. We describe this process saying a: \emph{Pub/Sub matching has taken place for unicast communication} where the RV coordinates the matching and the TM allocates a forwarding ID (FID) to the publishing NAP used to forward data. Network enabled mobility management using IP over ICN architecture including basic MN bootstrapping and handover is described in detail in our previous work \cite{mshawk2016} where interested readers are referred to.

%
Now consider that MN A moves such that will need to initiate a handover that will result, ultimately, in a handover to NAP C, although during the handover there may be some uncertainty about the correct target NAP.
The detailed sequence diagram for the handover process is illustrated in Fig. \ref{MBB2} through an example. This example assumes that the aforementioned Pub/Sub matching for unicast communication has already taken place, allowing MN A and MN B to communicate. In this example, the serving eNodeB (eNodeB A) continuously receives measurement reports from the MN (MN A) including signal quality metrics. It uses these metrics to decide if a potential handover is detected based on the quality metrics degrading to a certain level. Upon detecting a potential handover, eNodeB A starts a handover process that includes three phases (Handover Preparation, Execution and Completion). During handover preparation, the serving NAP (NAP A which is collocated with eNodeB A) sends a subscription message to NAP B (collocated with eNodeB B) to subscribe all the neighbouring NAPs identified by $/root/NAP(A)\_A$ of the handover namespace to receive packets destined for the IP address of MN A from MN B. Then, the local RV at NAP B matches the subscription with a previous publication to the IP address of MN A, and updates the forwarding route identifier (FID) to a multicast FID as there are now multiple destinations \emph{i.e.} the candidate handover eNodeBs/NAPs. Upon updating the FID, the NAP B starts sending RLC coded traffic and uses the updated multicast FID to multicast its traffic to NAP A and all its neighbouring NAPs so that traffic can reach MN A whichever neighbouring eNodeB/NAP it is going to move to. The RLC helps in dealing with packets' loss or delayed arrivals due to the handover process. When RLC is used, a receiver (user) requests coded packets that carry innovative information with respect to the packets already received (each coded packet contains information from multiple original uncoded source packets) and does not request a specific uncoded packet. This does not necessitate ARQ mechanisms which can further congest the core network. In addition, NAP B sends a multicasted message to NAP A and all its neighbouring NAPs with the Pub/Sub state. This state is stored at the participating NAPs until one of them takes ownership of the state after MN moves into its coverage area. The handover execution phase involves the Layer 2 link tear down at the previous serving NAP and the Layer 2 link up at the new serving NAP. When the link at the new serving NAP is up, then the handover completion phase starts. This phase consists of the session re-establishment where the new serving NAP (NAP C) re-initiates IP address establishment through DHCP, keeping the MN IP address unchanged. When the first packet from MN A to MN B through eNodeB C/NAP C is to be sent (or rather slightly earlier when link-layer connectivity is confirmed) NAP C sends a message to NAP B that indicates that it (NAP C) is publishing to /IP-Prefix/IP-B and and telling NAP B to implicitly subscribe to the scope of MN A's own IP address /IP-Prefix/IP-A (termed PubiSub). This triggers NAP B to stop using multicast with RLC and instead send unicast responses to MN A from NAP B to NAP C such that normal IP over ICN traffic can resume between MN A and MN B.
 
\section{Mobility Management Cost Analysis}
 
In this section, we analyse the mobility management cost of the proposed solution and compare it with the corresponding cost of Fast Handover in Proxy MPv6. In our analysis, for the sake of simplicity of description, we assume that only one end of the communication is mobile (MN) and that the corresponding node (CN) is static and that hence is not generating additional mobility signalling.
 
\subsection{Fast Handover in Proxy MPv6}

\begin{figure}[t]
    	\centering
	\includegraphics[width=0.9\columnwidth]{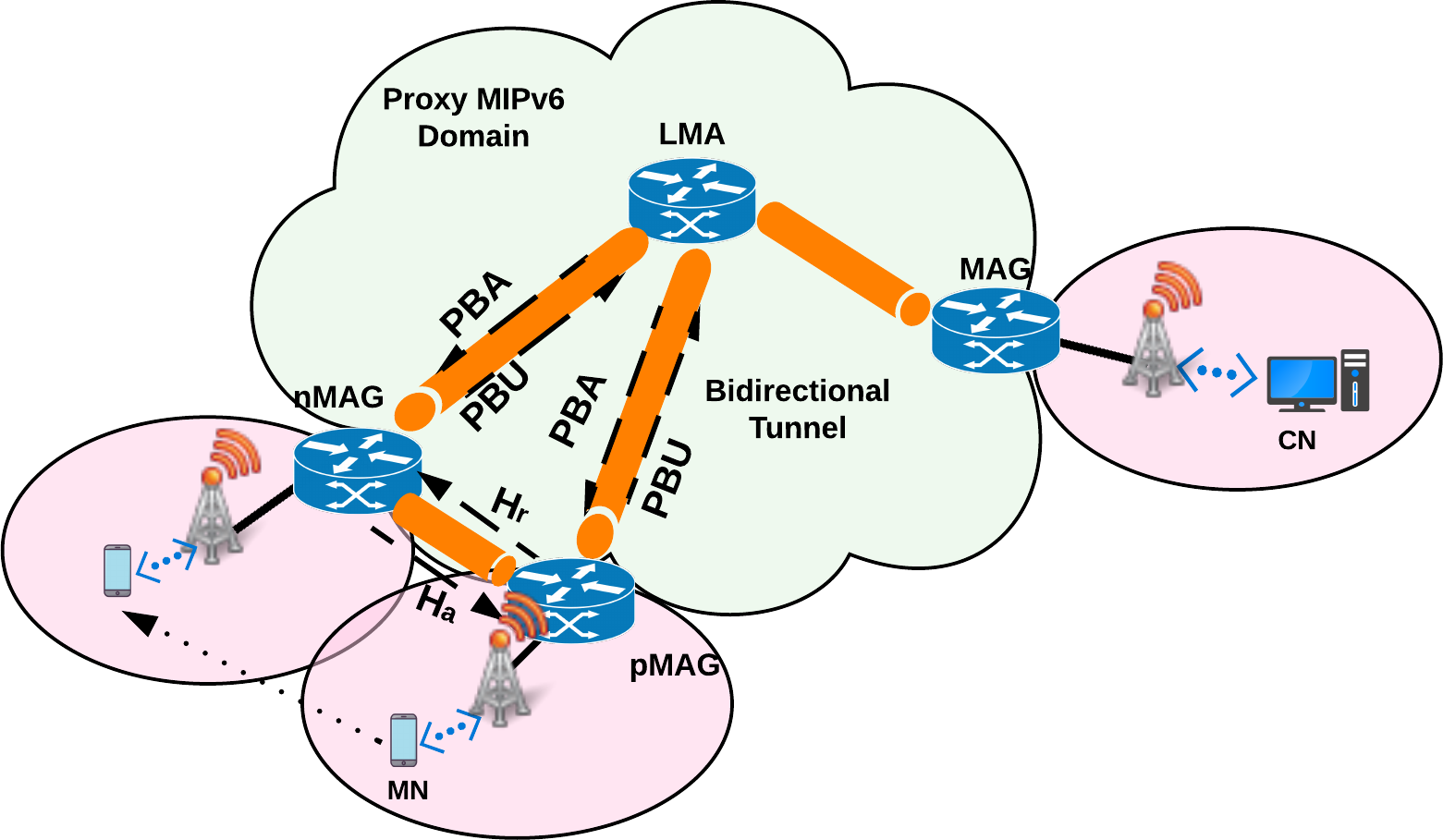}
    	\caption{Fast Handover in Proxy MPv6}
	\vspace{-5mm}
    	\label{FastHO2}
    \end{figure}

PFMIPv6 as depicted in Fig. \ref{FastHO2} performs a handover negotiation process for data forwarding from the previous Mobile Access Gateway (pMAG) to the next Mobile Access Gateway (nMAG) and for transferring the MN context before handover execution for the MN. If the Received Signal Strength Indicator (RSSI) of the MN with the serving eNodeB is less than a pre-defined threshold, the Handover-Initiate process is triggered and the following operations of PFMIPv6 handover are performed \cite{kim2010fast,kim2013performance}:
 
 \begin{enumerate}[leftmargin=*]
 \item The MN first scans its neighbouring access networks to find the eNodeB with the strongest RSSI. Then, the MN sends a L2 measurement report to its previous eNodeB containing its own and the next eNodeB IDs. 
 \item Upon receiving the report, the previous eNodeB indicates the MN's handover to the pMAG so that it can start to set up an IP-in-IP tunnel between the nMAG and itself. This is done by sending to the nMAG a Handover Initiation Request $H_{r}$ message containing MN's context information.
 \item As a reply to the $H_{r}$ message, the nMAG sends a Handover Acknowledgement message $H_{a}$ to the pMAG containing the MN's ID.
 \item Upon receiving the $H_{a}$ message, the pMAG starts forwarding the packets to the nMAG through the tunnel. The nMAG forwards the buffered packets to the next eNodeB once the MN is associated with the new access network.
\end{enumerate}

The previous scenario is for a predictive handover. If the MN hands over to the next eNodeB without sending a measurement report about the next eNodeB, to the previous eNodeB, a reactive handover process of PFMIPv6 is performed.
 
 \subsubsection{Mobility Signalling Cost}
 
The signalling cost consists of the PMIPv6 signalling cost \cite{lee2010cost} represented by the proxy binding update $PBU$ and proxy binding acknowledgement $PBA$ at the previous MAG and next MAG sent towards the Local Mobility Anchor (LMA), in addition to the signalling overhead caused by setting up a tunnel between the pMAG and nMAG for fast handover support. The total signalling cost for successful predictive and reactive PFMIPv6 handovers is given by:
\begin{eqnarray}
\label{emc13}
 	\Upsilon & = & (1+P)\big\{h_{p,m}(|PBU| + |PBA|) \\ \nonumber
		   & + &  h_{n,m}(|PBU| + |PBA|) + h_{p,n}(|H_{r}| + |H_{a}|)\big\},
\end{eqnarray}
where $P$ is the handover failure probability, $h_{p,m}$ stands for the number of hops between pMAG and the LMA, $h_{n,m}$ is the number of hops between the nMAG and the LMA, and $h_{p,n}$ denotes the number of hops between pMAG and the nMAG. In \eqref{emc13}, $|H_{r}|$ represents the message size \footnote{In this paper, $|x|$ denotes the length of message $x$.}  in \textit{bytes} of the handover initiation request sent from pMAG to the nMAG during predictive handover or from nMAG to the pMAG during reactive handover and $|H_{a}|$ is the message size in \textit{bytes} of the handover acknowledgement sent from nMAG to the pMAG during predictive handover or from pMAG to the nMAG during reactive handover. By setting $|PB|$ = $|PBU|$ + $|PBA|$, \eqref{emc13} is rewritten as:
\begin{eqnarray}
\label{emc13new}
 	\Upsilon & = &(1+P)\big\{|PB|(h_{p,m} + h_{n,m}) \\ \nonumber
 	& + & h_{p,n}(|H_{r}| + |H_{a}|)\big\}
\end{eqnarray}

\subsubsection{Mobility Packet Delivery Cost}
The packet delivery cost $\Lambda$ represents the packet delivery overhead needed to support fast mobility. It is calculated as the product of the average packet arrival rate in \textit{packets/sec}, packet size in \textit{bytes} and the hop distance. The packet delivery cost for PFMIPv6 is measured in \textit{Bytes$\times$Hops/Sec} and is expressed as:
  \begin{equation}\label{emc1}
  	\Lambda = (1+P)RO,
  \end{equation}
where $R$ and $O$ are the average packet arrival rate, and the direct path packet cost in PFMIPv6, respectively. $O$ can be calculated as
 %
 \begin{eqnarray}
\label{emc1}
	O & = & (h_{c,m} + h_{m,p} + h_{p,n}) ( \varphi + \zeta ),
\end{eqnarray}
where $h_{c,m}$, $h_{m,p}$, and $h_{p,n}$ denote the number of hops between the corresponding node (CN) and the LMA, the number of hops between the LMA and the pMAG, and the number of hops between the pMAG and the nMAG, respectively. Finally, the parameter $\zeta$ is the average data packet length and $\varphi$ represents the tunnelling overhead in \textit{bytes}. 
 %
 \subsection{Seamless Handover in IP over ICN}
 
For seamless Handover in IP over ICN, as illustrated in Fig. \ref{MBB2}, a handover preparation process is performed that includes multicasting RLC coded traffic from the corresponding NAP to the handover neighbourhood and also multicasting the MN Pub/Sub state before the handover execution. If the RSSI of the MN with the serving access network is less than a pre-defined threshold, the handover preparation process for the MN is triggered and the following operations are performed:
  
  \begin{enumerate}[leftmargin=*]
  \item The NAP on the previous link (NAP A) signals the corresponding NAP B by sending a group subscription message $\ell_{s}$ to the scope of NAP A's neighbouring NAP's represented by a hashed function of NAP A's ID $/root/NAP(A)\_A$. 
  \item NAP B matches the subscription with a previous publication to the IP address of MN A, it then updates the previous FID to a multicast FID and uses the new FID to send a multicast stream of RLC coded traffic to all neighbouring NAP's of NAP A in addition to NAP A itself. 
  \item NAP B sends a multicasted state update message $\ell_{u}$ to NAP A and all its neighbouring NAP's that includes the Pub/Sub state for MN A to be stored at the participating NAPs. This message is used when MN A moves into one of the participating NAPs coverage area.
  \item After MN A establishes Layer 2 connectivity with NAP C and IP address allocation, NAP C receives the first IP packet destined to the CN at NAP B, and looks up locally the appropriate FID to reach NAP B and uses it to send a PubiSub message $\ell_{i}$ to NAP B. This message includes the first data packet from the MN to the CN in addition to an implicit subscription to MN A's own IP address scope.
  \item The PubiSub message triggers NAP B to utilize its local Rendezvous in order to maintain a match Pub/Sub relation for the mentioned scope, looks up its local database for the appropriate FID to reach NAP C and use it to start publishing information to the identified subscriber.
  \end{enumerate} 
At this point MN A and CN B can commence sending and receiving data payload messages of size $\zeta$.

\subsubsection{Mobility Signalling Cost} 

The mobility signalling cost $\Upsilon^\prime$ is the size of the signalling messages in \textit{bytes} multiplied by the number of hops. Therefore, the introduced signalling overhead for seamless handover in IP over ICN is computed as follows:
 \begin{equation}\label{emc1a}
 	\begin{split}
 		\Upsilon^\prime = \big\{h_{a,b} |\ell_{s}| + h_{c,b} |\ell_{i}| + h_{b,j} |\ell_{u}| + \sum_{n=1}^{N} h_{j,n} |\ell_{u}|\big\},
 	\end{split}
 \end{equation}
 where $h_{a,b}$ is the number of hops between the previous NAP A and the corresponding NAP B, $h_{c,b}$ is the number of hops between the next NAP C and the corresponding NAP B, $h_{b,j}$ is the number of hops between the corresponding NAP B and the multicast route fan out node $j$. Finally, $h_{j,n}$ is the number of hops between the multicast route fan out node $j$ and neighbouring NAP $n$.
 
\subsubsection{Mobility Packet Delivery Cost} 
 
The packet delivery cost $\Lambda^\prime$ is mainly used to investigate the packet delivery overhead of the RLC coded stream used to support seamless mobility. The packet delivery cost is calculated as the product of average packet arrival rate in \textit{packets/sec}, the packet size in \textit{bytes} and the hop distance. The packet delivery cost for IP over ICN in \textit{Bytes$\times$Hops/Sec} is expressed as follows:
 \begin{equation}\label{emc1b}
 	\Lambda^\prime =  R^{'}O^{'}, 
 \end{equation}
where  $O^{'}$ is the direct path overhead for a RLC coded packet in an IP over ICN network and $R^{'}$ denotes the average packet arrival rate of RLC coded packets for the file transmitted over the network during handover, and is calculated as $R^{'}=R(1+\epsilon)$. $\epsilon$ stands for the coding overhead and is equal to \textit{\( \frac{K}{N}-1\)} with $K$ representing the number of received packets and $N$ the number of source packets. Since the code structure of the employed RLC codes is random, each packet should contain a header with the coding coefficients that describe the coding operations that were performed to generate the coded packet. The size of this header depends on the number of source packets and the Galois field where the operations are performed. We compress this header in 2 bytes by following the approach used in Raptor codes \cite{shokrollahi2006raptor}, \textit{i.e.}, we describe the coding coefficients in each packet only with the seed of the pseudorandom generator used for producing the coding coefficients.

We can obtain $O^{'}$ in \eqref{emc1b} as follows:
 \begin{equation}\label{emc1b}
 	O^{'} = h_{b,j}( \varphi^{\prime} + \zeta ) + \sum_{n=1}^{N} h_{j,n} (\varphi^{\prime} + \zeta),
 \end{equation}
where $h_{b,j}$ is the number of hops between NAP B where the CN is attached and the multicast route fan out node $j$, $h_{j,n}$ is the number of hops between the multicast route fan out node $j$ and neighbouring NAP $n$. Finally, $\varphi^{\prime}$ represents the size of the ICN payload packet header. 

The average number of RLC packets $K$ that should be sent in order to recover the $N$ source packets according to \cite{trullols2011exact} is:
 \begin{equation}\label{emc2}
  	K =  \sum_{k=N}^{\infty} k \cdot P_{d}(k,N),
\end{equation}
where $P_{d}$ denotes the probability that the receiver node has obtained $N$ linear independent packets over the $K$ transmitted packets. It is calculated as:
  \begin{equation}\label{emc2}
  	P_{d}(K,N) = 
  	\begin{cases}
  		0  &\text{, if $K < N$}\\  
     \prod_{j=0}^{N-1} 1-\frac{1}{q^{K-j}} &\text{, if $K \geq N$}	 
  	\end{cases},
  \end{equation} 
 where $q$ is the size of the employed Galois Field. 
 \begin{table}[t]
 	\caption{List of mobility messages and their sizes}
 	\label{tab:3}
 	\centering
 	\begin{tabular}{|c||c||c|}
 		\hline
 		Notation    &   Description   & Size\\
 		\hline
 		$PBU$ & Proxy binding update & 76 Bytes \cite{lee2010cost}\\
 		\hline
 		$PBA$ & Proxy binding acknowledgement & 76 Bytes \cite{lee2010cost}\\
 		\hline
 		$H_{r}$ & Handover Initiation Request  & 104 Bytes \cite{kim2013performance}\\
 		\hline
 		$H_{a}$ & Handover Acknowledgement  & 168 Bytes  \cite{kim2013performance}\\
 		\hline
 		$\varphi$ & Proxy MIPv6 tunnelling header & 40 Bytes \cite{lee2010cost}\\
 		\hline
 		$\zeta$ & Average payload length & 1024 Bytes\\
 		\hline	
 		$\ell_{u}$   & multicasted state update message& 102 Bytes \\
 		\hline
 		$\ell_{s}$   & group subscription message & 102 Bytes \\
 		\hline
 		$\ell_{i}$   & Publish with Implicit Subscription & 166 Bytes \\
 		& message (PubiSub)&  \\
 		\hline
 		$\varphi^{\prime}$   & ICN payload packet header & 96 Bytes \\
 		\hline
 		\noalign{\smallskip}
 	\end{tabular}
	\vspace{-5mm}
 \end{table} 
\section{Simulation and Cost Evaluation}
To evaluate the performance of the proposed seamless handover solution in IP over ICN networks and the FPMIPv6, we built a discrete time event simulation in $R$. 
We considered an operator like network topology of 71 nodes consisting of 8 core forwarder nodes and 60 eNodeBs. MAGs and NAPs in both FPMIPv6 and IP over ICN scenarios have been assumed to be collocated with the eNodeBs. We assume that MAGs and NAPs have a circular coverage area of radius 500 m. For our simulations, the same central node was used to represent the LMA and TM/RV in order to ensure accurate cost comparisons. A random walk mobility model has been used to capture user mobility at 70 miles/hour. Users' initial locations are decided using a uniform random distribution. In our traffic model, we assume that all the users in the network are exchanging video data with an arrival rate of 1 Mbps following a Poisson distribution. All the coding operations are performed in $GF(2^4)$. Table \ref{tab:3} includes a summary of the mobility messages and their relevant sizes for both evaluated solutions. In the case of $\ell_{u}, \ell_{s}, \ell_{i}$ and $\varphi^{\prime}$ these assume ICN FIDs and scope/ID lengths of 256 bits each and a single scope and ID for the IP naming.
\vspace{-1.5mm}
\subsection{Performance Evaluation with respect to Handover Failure Rates}
\begin{figure}[t]
	\centering
	\includegraphics[width=\columnwidth,height=2.8in]{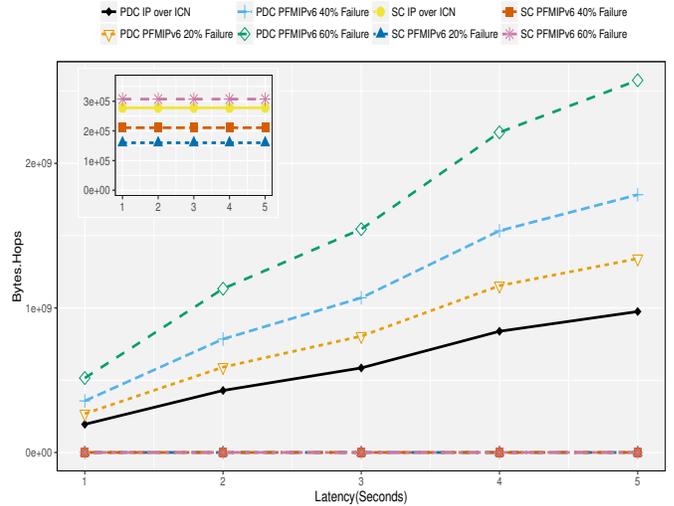}
	\caption{Handover Packet Delivery Cost (PDC) and Signalling Cost (SC) for PFMIPv6 and IP over ICN with respect to Different Handover Latency times and Handover Failure Rates.}
	\vspace{-5mm}
	\label{PDC2}
\end{figure}

In Fig. \ref{PDC2}, we compare the handover performance of both evaluated solutions in terms of packet delivery and signalling cost for a total of 100 handovers according to different handover failure rates (ranging between 20\% and 60\%) for PFMIPv6 as shown in Section 3. This figure also illustrates the impact of different handover latency times on the resulting costs. From Fig. \ref{PDC2}, we can see that as the handover failure rate for PFMIPv6 increases from 20 to 60\%, the total packet delivery cost doubles from approximately \num{2.5e 08} to \num{5e 08} \textit{Bytes.Hops} when the evaluated handover latency was only one second. Moreover, as the handover latency increases, the total packet delivery cost for both PFMIPv6 and IP over ICN increases accordingly. This comparison clearly reveals the benefit of using RLC coded traffic during handover in IP over ICN networks where handover failures are eliminated due to the availability of the handover traffic at all the eNodeBs within the handover neighbourhood and also due to the characteristics of RLC coded traffic that makes reassembling asynchronous randomly linear coded packets possible. From Fig. \ref{PDC2}, we can observe that these benefits come with a higher signalling cost in the ICN core, although the difference in the signalling cost tends to dissipate as the handover failure rates increase for PFMIPv6 reaching about \num{3e 05} \textit{Bytes.Hops} with 60\% handover failure compared to \num{2.8e 05} \textit{Bytes.Hops} for IP over ICN.

\begin{figure}[t]
	\centering
	\includegraphics[width=\columnwidth,height=2.8in]{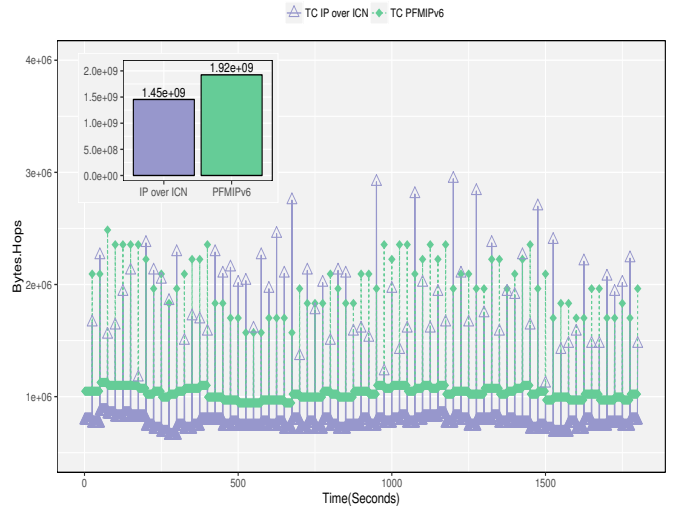}
	\caption{Total Cost (TC) including Packet Delivery Cost and Signalling Cost for PFMIPv6 vs. IP over ICN for MN's in Handover + Non Handover Mode.}
	\vspace{-5mm}
	\label{PDC3}
\end{figure}

\begin{figure}[t]
	\centering
	\includegraphics[width=\columnwidth]{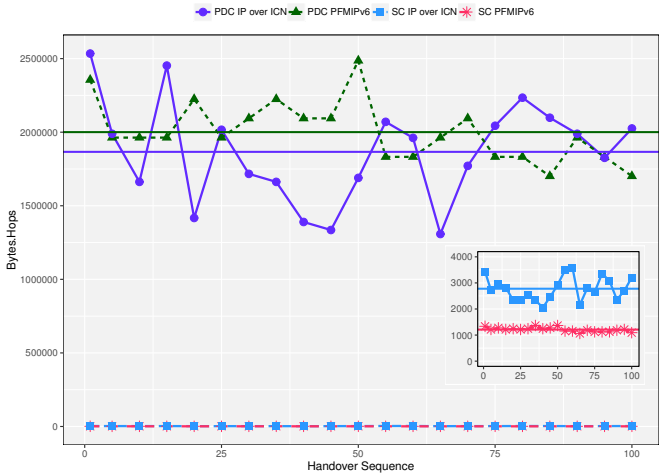}
	\caption{Averaged Packet Delivery Cost (PDC) and Signalling Cost (SC) for PFMIPv6 vs. IP over ICN with respect to 100 sequent Handover's.}
	\vspace{-5mm}
	\label{PDC1}
\end{figure}

\subsection{Performance evaluation with respect to MNs in Handover + Non Handover mode}

Fig. \ref{PDC3} shows the results of a simulation run of 1800 seconds for both PFMIPv6 and seamless handover in IP over ICN networks where 35 MNs are moving within a PFMIPv6 or IP over ICN domain of 70 nodes previously described. The MNs incur both handover and non handover traffic and signalling according to their location and direction at each point in time. It is assumed that handover latency is 1 second on average. It is also worthy to note that no handover failure has been assumed in this simulation experiment for PFMIPv6 in order to compare the evaluated solutions with no dependencies despite that this is impractical due to the reasons highlighted previously in the paper. In Fig. \ref{PDC3}, handover costs (including packet delivery + signalling costs) are indicated by the sparks of traffic shown in the graph at relevant points in time while non handover costs (also in terms of packet delivery + signalling costs) can be clearly differentiated by their steady pattern. In total, IP over ICN incurs a cost of \num{1.45e 09} compared to \num{1.92e 09} for PFMIPv6. Therefore, the figure clearly shows that using an ICN core with the help of RLC coding to facilitate IP mobility with better QoS (no packet loss and no HO failure) can be achieved with lower costs than PFMIPv6 in average and in total. 

\subsection{Performance evaluation with respect to Handover costs}

We finally investigate the performance of all the schemes under comparison with respect to the handover cost. Fig. \ref{PDC1} shows the average and mean packet delivery and signalling costs for 10 MNs for 100 sequent handovers. We observe that PFMIPv6 incurs a slightly higher mean packet delivery cost of approximately \num{2e 06} Bytes.Hops compared to \num{1.85e 06} \textit{Bytes.Hops} for IP over ICN. On the other hand, IP over ICN incurs a higher signalling cost of approximately \num{2.75e 03} \textit{Bytes.Hops} compared to \num{1.25e 03} \textit{Bytes.Hops} for PFMIPv6. This is due to the fact that IP over ICN uses source routing and therefore imposes more signalling messages to secure delivery path trees to the traffic source during mobility. Overall, our scheme outperforms PFMIPv6 significantly in terms of the total cost, as the packet delivery cost is the dominant cost.

\section{Conclusion}

In this paper, we follow a novel random linear coding approach for seamless IP handover with minimum disruption. We have shown that seamless handover with no packet loss or handover failure is achieved with lower cost than that of existing state-of-the-art solutions such as Fast Handover for Proxy MIPv6 (PFMIPv6). The gains come from the use of RLC coding which removes the need for resuming packet transmissions when a handover occurs and packet synchronization. Also, the use of ICN in the core network ensures optimal shortest path packet delivery rather than the traffic anchoring and tunnelling methods used by the currently available IP solutions that lead to using non-optimum roots for packet delivery. Further, there are significant gains in high mobility environments despite the small increase in the mobility signalling cost of IP over ICN. Moreover, IP over ICN has been shown to provide better QoS than current LTE handover solutions especially in terms of handover failure rates which are inevitable in existing solutions with a minimum of 20\% failure.





\bibliographystyle{IEEEtran}
\bibliography{ref_nikosedited}
\end{document}